\documentclass[prb, secnumarabic, twocolumn, graphics, floatfix, superscriptaddress, tightenlines, longbibliography, aps]{revtex4-1}
\usepackage[dvips]{graphicx}
\usepackage{dcolumn}
\usepackage{bm}
\usepackage{color}
\usepackage{array}
\usepackage{upgreek}
\usepackage{amsmath}
\newcolumntype{L}[1]{>{\raggedright\let\newline\\\arraybackslash\hspace{0pt}}m{#1}}
\newcolumntype{C}[1]{>{\centering\let\newline\\\arraybackslash\hspace{0pt}}m{#1}}
\newcolumntype{R}[1]{>{\raggedleft\let\newline\\\arraybackslash\hspace{0pt}}m{#1}}

\usepackage[unicode=true,pdfusetitle,
 bookmarks=true,bookmarksnumbered=true,bookmarksopen=false,
 breaklinks=true,pdfborder={0 0 0},pdfborderstyle={},backref=false,colorlinks=true]
 {hyperref}
\hypersetup{
 pdfborderstyle={},pdfborderstyle={},pdfborderstyle={},pdfborderstyle={}}
\usepackage{breakurl}

\begin{document}

\title{Single-beam resonant spin amplification of electrons interacting with nuclei in a GaAs/(Al,Ga)As quantum well}

\author{M.~Kotur}
\affiliation{Experimentelle Physik 2, Technische Universit\"at Dortmund, D-44221 Dortmund, Germany}
\affiliation{Ioffe Institute, Russian Academy of Sciences, 194021 St.Petersburg, Russia}

\author{F.~Saeed}
\affiliation{Experimentelle Physik 2, Technische Universit\"at Dortmund, D-44221 Dortmund, Germany}
\affiliation{Centre for Micro and Nano Devices, Department of Physics, COMSATS University Islamabad, Park Road, Islamabad 44000, Pakistan}

\author{R.~W.~Mocek}
\affiliation{Experimentelle Physik 3, Technische Universit\"at Dortmund, D-44221 Dortmund, Germany}

\author{V.~L.~Korenev}
\affiliation{Ioffe Institute, Russian Academy of Sciences, 194021 St.Petersburg, Russia}

\author{I.~A.~Akimov}
\affiliation{Experimentelle Physik 2, Technische Universit\"at Dortmund, D-44221 Dortmund, Germany}
\affiliation{Ioffe Institute, Russian Academy of Sciences, 194021 St.Petersburg, Russia}

\author{A.~S.~Bhatti}
\affiliation{Centre for Micro and Nano Devices, Department of Physics, COMSATS University Islamabad, Park Road, Islamabad 44000, Pakistan}

\author{D. R. Yakovlev}
\affiliation{Experimentelle Physik 2, Technische Universit\"at Dortmund, D-44221 Dortmund, Germany}
\affiliation{Ioffe Institute, Russian Academy of Sciences, 194021 St.Petersburg, Russia}

\author{D.~Suter}
\affiliation{Experimentelle Physik 3, Technische Universit\"at Dortmund, D-44221 Dortmund, Germany}

\author{M.~Bayer}
\affiliation{Experimentelle Physik 2, Technische Universit\"at Dortmund, D-44221 Dortmund, Germany}
\affiliation{Ioffe Institute, Russian Academy of Sciences, 194021 St.Petersburg, Russia}

\begin{abstract}
The dynamic polarization of nuclear spins interacting with resident electrons under resonant excitation of trions is studied in a nominally undoped GaAs/(Al,Ga)As quantum well. Unlike in common time-resolved pump-probe techniques, we used a single beam approach where the excitation light is modulated between the circular and linear polarization states. The time-integrated intensity of the excitation laser reflected from the sample surface, proportional to the optical generation rate and changes due to the pumping of the resident electrons, is detected. Polarized electrons on the other hand transfer their spin to the lattice nuclei via the hyperfine interaction. Exciting the sample with a train of pulses in an external magnetic field leads to resonant spin amplification observed when the Larmor precession frequency is synchronized with the laser pulse repetition rate. Build-up of the nuclear spin polarization causes a shifting of the RSA peaks since the resulting nuclear field alters the strength of the external magnetic field experienced by the electrons. It was established that the nuclear spin polarization time $T_1$ is temperature dependent and owing to the electron localization at lower temperatures becomes shorter. "Locking" of the nuclear field manifested as the limited by the strength of the external field growth of the nuclear field, that is related to the anisotropy of the electron $g$-factor, was observed. The $g$-factor ratio between the in plane $g_{\parallel}$ and out-of-plane $g_{\bot}$ components was estimated to be $g_{\bot}/g_{\parallel}=1.3$.

\end{abstract}

\pacs{} \maketitle

%
\section{Introduction}
The idea of spin utilization for information processing and storage induced comprehensive research in recent years \cite{vzutic2004spintronics}. Nuclear spins are strong candidates for long-term memory storage owing to the weak coupling of the nuclear spin system (NSS) to the lattice resulting in a long spin lifetime \cite{dyakonov2017spin, meier2012optical}. Therefore, understanding the processes of nuclear spin polarization and relaxation are important both from the fundamental and the technological point of view. One of the most common techniques used to study nuclear spin polarization is the optical orientation method in oblique external magnetic field where circularly polarized light from a continuous wave (CW) laser is used to dynamically polarize the nuclei via the hyperfine interaction with optically polarized electrons. The state of the nuclear spin system (NSS) is then detected by analyzing the polarization degree of the photoluminescence (PL) which is proportional to the mean electron spin \cite{meier2012optical}.

Time-resolved polarization spectroscopy based on pump-probe Faraday/Kerr rotation techniques with a pulsed laser source gives the opportunity to monitor the evolution of the mean electron spin in time and to directly measure the Larmor precession frequency of the electrons \cite{awschalom2013semiconductor}. Knowing the Larmor frequency allows one to determine the electron $g$-factor and spin relaxation time. Since the electron Larmor precession frequency changes as a result of the dynamic nuclear polarization (DNP) in an oblique external magnetic field, it is possible to determine the strength of the nuclear magnetic field $B_N$ [\onlinecite{malinowski2000dynamic}]. The typical experimental routine consists of optical polarization of the nuclear spin using a circularly polarized train of pulses. The effect of the nuclear field is then monitored by measuring the oscillating projection of the electron spin onto the optical axis using the Faraday or Kerr rotation of the linearly polarized probe pulse which is time delayed in relation to the pump pulse. Even though time resolved pump-probe Faraday/Kerr rotation is a powerful and versatile method, its shortcomings are that it requires a mechanical delay line, a complex arrangement of the pump and probe beams and a polarization sensitive differential detection \citep{dyakonov2017spin}.

Herein we present a simpler approach used to study the nuclear spin dynamics in a nominally undoped GaAs/(Al,Ga)As quantum well (QW). A pulsed laser beam, whose photon energy is set in resonance with the transition of the negativley charged exciton (trion), is used for optical pumping of the resident electrons, which in turn polarize the nuclei by means of the hyperfine interaction. The state of the NSS is detected via the Hanle effect, i.e. the depolarization of the electron spin in a transverse magnetic field, which is modified by the presence of the Overhauser field of the spin-polarized nuclei. To that end, an external magnetic field is applied in oblique, but nearly Voigt geometry. The electron spin polarization is monitored via the intensity of the excitation laser reflected from the sample surface. Since we used a laser source with relatively high repetition rate $f$ ($f\gg\tau_s^{-1}$, where $\tau_s$ is the spin relaxation time), resonant spin amplification (RSA) is recorded for each external field when the resonance condition $\omega_L=2\pi nf$ is satisfied, where $n$ is an integer. Therefore, our approach can be described in terms of optical pumping of the electron spin into a state stationary in the frame rotating with a frequency corresponding to $2\pi nf$. The first experiment using similar scheme was conducted on atoms more than 50 years ago \cite{bell1961optically} and it was recently applied to a semiconductor CdTe/(Cd, Mg)Te QW structure \cite{saeed2018single}.

The advantage of the pulse over CW excitation is that it allows observing the effects of the nuclear field even in relatively strong magnetic fields. In an oblique geometry, when the nuclear spin polarization is studied by the Hanle effect, the difference in the degree of polarization of luminescence is observed due to the increase of the nuclear field \citep{meier2012optical}. Nonetheless, if the external field is strong enough to completely depolarize the luminescence, effects associated to the rise of the nuclear field remain unnoticed, when the two fields enhance each other. On the other hand, pulsed excitation leads to a periodic repetition of the RSA peaks, which allowed us to study nuclear spin dynamics for both signs of the external magnetic field. In the first case, when the vectors of the external and the nuclear field have the same direction, a shifting of the RSA peaks was noticed due to the build-up of the nuclear field, which made possible to extract the nuclear spin polarization time $T_1$ as well as the value of the stationary nuclear field for three different temperatures. With the change of the external field sign, it becomes anti-parallel to the nuclear field. In this case, it was noticed that the strength of the nuclear field is limited by the strength of the external field. A similar behavior was reported using the Hanle effect, where, in the stationary state, it is noticed that the nuclear field reaches the value of the external field whereupon it continues to follow the strength of the external magnetic field with its further increase \cite{snelling1991magnetic, kalevich1992anisotropy}. Such behavior was associated with the anisotropy of the electron $g$-factor. Here, we addressed the problem of nuclear spin "locking" in a dynamic approach, where we measured the build-up of nuclear field in various external magnetic fields. It was observed that the nuclear field rises from zero until full compensation of the external field is reached without further increase.

\section{Sample and experimental setup}
\begin{figure*}[t]
\center{\includegraphics[width=0.7\linewidth]{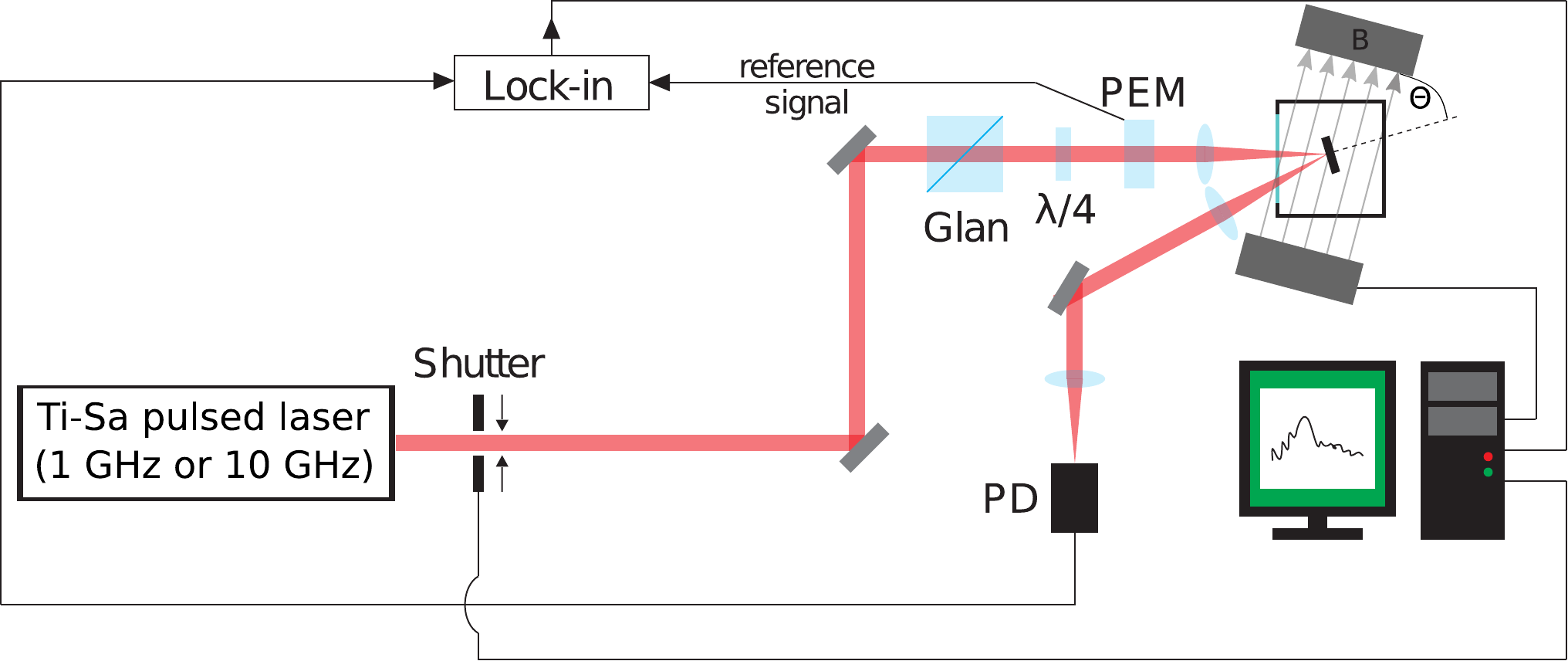} }
\caption{Schematic representation of the experimental setup. Both mechanical shutter and power supply for the magnetic field were controlled by the computer which allowed for a simultaneous change of the shutter status (open/closed) and the strength of the field. PEM and PD denote photoelastic modulator and photodiode.}
\label{fig:setup}
\end{figure*}

In this work we used a sample composed of 13 nominally undoped $\textnormal{GaAs/(Al}_{0.35}\textnormal{Ga}_{0.65})\textnormal{As}$ QWs with thicknesess ranging from 2.8 up to 39.3 nm, separated by 30.9 nm thick barriers and grown by molecular-beam epitaxy on a Te-doped GaAs substrate. The sample was previously investigated in Refs. [\onlinecite{mocek2017high},\onlinecite{eickhoff2002coupling}]. The PL spectrum of the $d=19.7\,\textnormal{nm}$ QW of interest consists of two peaks at 812.1 nm and 812.7 nm which are attributed to the emission from the exciton ($X$) and trion ($X^-$) complexes, respectively \citep{mocek2017high}. Although our sample is nominally undoped, the presence of the trion line indicates the presence of low-density ($n_e\leq10^{10}\;\textnormal{cm}^{-2}$) resident electrons. The sample was placed in a liquid helium bath cryostat, where the temperature $T$ was varied from 1.8 up to 10 K. The external magnetic field $B$, provided by an electromagnet, was applied in oblique geometry ($75\pm5^{\circ}$ in relation to the QW growth axis) in order to allow for both dynamic nuclear polarization ($B_{\parallel}$) and detection due to the Hanle effect ($B_{\bot}$). The suggested technique was used with two different tunable Ti-Sapphire lasers, self mode-locked femtosecond oscillators with: (i) repetition rate of $f=1\;\textnormal{GHz}$, spectral width of 30 nm and pulse duration of 50 fs; or (ii) repetition rate of $f=10\;\textnormal{GHz}$, spectral width of 20 nm and pulse duration of 50 fs. The laser spectral width was reduced by a pulse shaper down to $\sim1\;\textnormal{nm}$ resulting in a longer pulse duration of several ps. The lasers with high repetition rate were used in order to achieve a reasonable distance ($\omega_L=2\pi nf$) between the RSA peaks. Reducing the pulse repetition frequency would result in a convergence of the RSA peaks making it more difficult to follow the dynamics of the nuclear spin. The excitation energy was set to the trion resonance and the reflected light intensity was measured with a photodiode PD (Fig. \ref{fig:setup}).

\begin{figure}
\center{\includegraphics[width=1\linewidth]{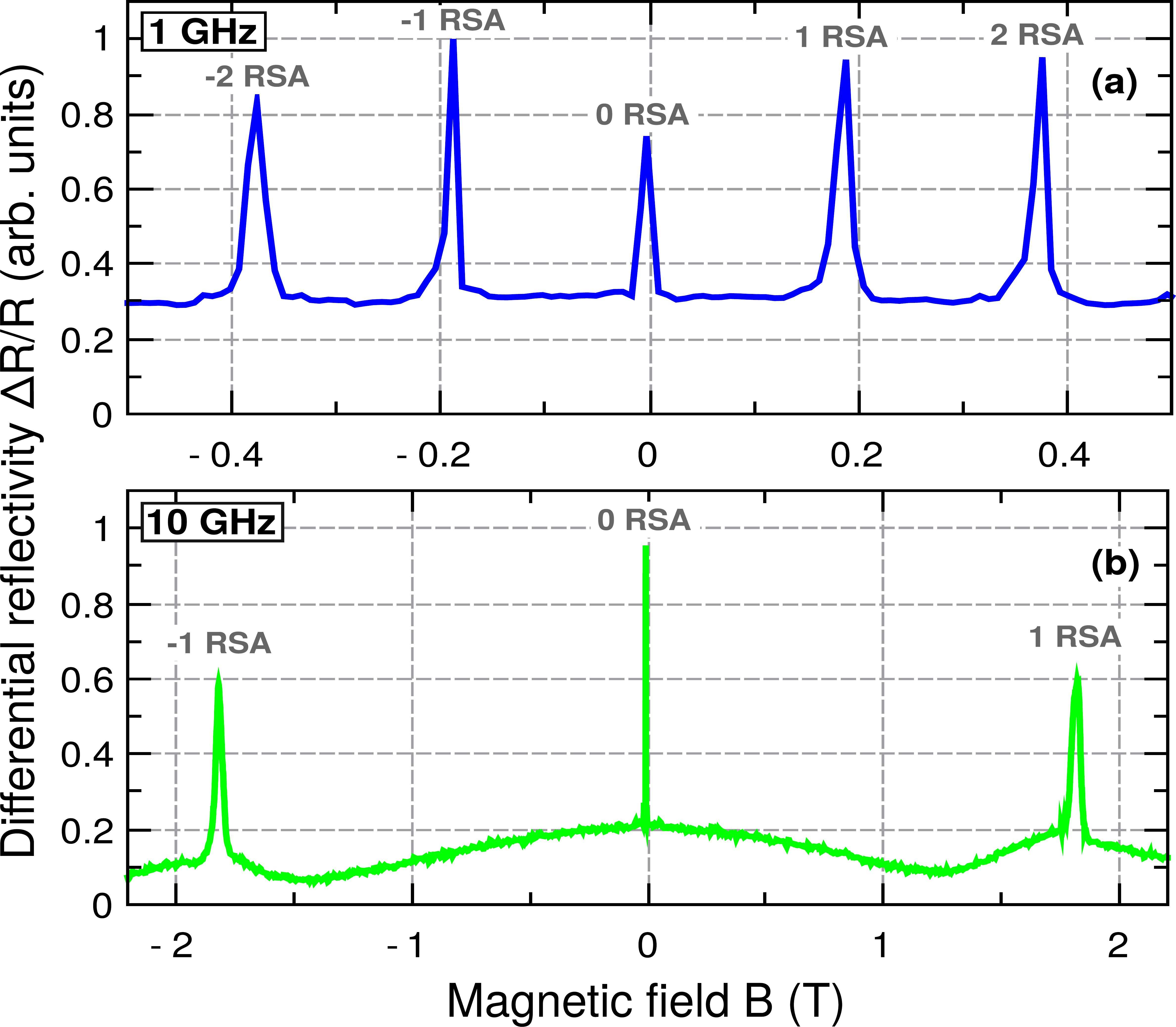}} 
\caption{Magnetic field dependence of the differential reflectivity for resonant excitation of the trion with the 1 GHz (a) and 10 GHz (b) pulsed laser, measured at $T = 1.8\;\textnormal{K}$.}
\label{fig:RSA}
\end{figure}

Resonant excitation of the negatively charged exciton (trion) with circularly polarized light leads to spin orientation of resident electrons \cite{zhukov2007spin, dyakonov2017spin}. Linearly polarized light impinging at 45$^\circ$ to the optical axis of a photoelastic modulator (PEM) is modulated at the PEM operating frequency $\nu$ to be in phase between the parallel and orthogonal component during time $t$, i.e. $\phi=\frac{\pi}{2}\textnormal{sin}(2\pi\nu t)$ and the polarization of the exciting beam changes from linear ($\pi$) to circular ($\sigma^+$ or $\sigma^-$) and back to linear ($\pi$). Modulation of the excitation light with the PEM at a frequency of $\nu=50 \;\textnormal{kHz}$ allows us to detect the difference in intensity of the reflected light induced by the optical orientation of resident electrons achieved during $\sigma^+/\sigma^-$ excitation which perishes as the light becomes linearly polarized. In typical resonant spin amplification (RSA) experiments \cite{awschalom2013semiconductor, yugova2012coherent}, the change of electron spin polarization with magnetic field is measured at a constant delay between the pump and probe beams. For magnetic fields at which the frequencies of the pulse repetition and the spin precession are strictly resonant, an enhancement of the spin polarization is observed because of the phase synchronization between the spin injection and the precessing spin polarization \cite{kikkawa1998resonant}. When the spin polarized resident electrons are exposed to a transverse external magnetic field, under the weak photo-excitation, the equation of motion for the spin density $\mathbf{S}$ consists of three terms: (i) generation of the spin polarization by the optical pumping with circularly polarized light, (ii) spin precession induced by the external magnetic field and (iii) spin relaxation \cite{astakhov2008time}
\begin{equation}
\frac{d\mathbf{S}}{dt}=\mathbf{P}+\bm{\upomega}_L \times\mathbf{S}-\frac{\mathbf{S}}{\tau_s},
\label{eq:motion}
\end{equation}
where $\bm{\upomega}_L$ is the electron Larmor frequency and $\tau_s$ is the spin relaxation time of the resident electrons. $\mathbf{P}$ is the spin pumping term proportional to the optical generation rate $G=(G^++G^-)/2$, where the two contributions $G^+$ and $G^-$ are related to the right-hand ($\sigma^+$) and left-hand ($\sigma^-$) polarized excitation. The optical generation rate depends on the laser intensity $I$ and corresponds to the number of electron-hole pairs generated by light in a unit volume per unit time. In case of CW excitation, the steady state solution of Eq. \eqref{eq:motion} has the form commonly used to express the Hanle effect
\begin{equation}
S_z=\frac{P\tau_s}{1+\omega_L^2\tau_s^2}.
\label{eq:CWExcitation}
\end{equation}
Switching to pulsed excitation results in the optical generation rate becoming time dependent $G(t)=G_0\sum \textnormal{exp}(i2\pi n ft)$. Here, $G_0$ is proportional to the laser intensity $I$ integrated over time. In this context, the steady state solution of Eq. \eqref{eq:motion} is \cite{saeed2018single}
\begin{equation}
S_{zn}=\frac{1}{2}S_{z0}(\omega_L-2\pi nf).
\label{eq:PulsedExcitation}
\end{equation}
Thus, for excitation with a pulsed laser source a sequence of RSA peaks should be observed when for higher harmonics $|n|>0$ the Larmor precession frequency $\omega_L$ is replaced with $\omega_L-2\pi nf$ in Eq. \eqref{eq:CWExcitation}. The difference in reflected light intensity for circularly ($\sigma^+$/$\sigma^-$) and linearly polarized ($\pi$) beam is $\Delta R \propto I|S_{zn}|$ \cite{saeed2018single}. The magnetic field dependence of the differential reflectivity for resonant excitation of the trion with pulsed 1 GHz and 10 GHz lasers is shown in Fig. \ref{fig:RSA}. Indeed, due to the pulsed excitation, beside the zero field RSA peak, we observe RSA peaks repeating every 185.1 mT for the 1 GHz and 1.8 T for the 10 GHz laser. From the resonance condition $\omega_L=2\pi nf$, where $\omega_L=g\mu_BB/\hbar$, it is possible to calculate the electron $g$-factor: $|g|=0.38$. Knowing the $g$-factor value allows determination of the magnetic fields where phase synchronization condition is fulfilled for a given laser repetition frequency $f$. The difference in width of the peaks is almost negligible for the 1 GHz excitation in the range of applied fields. When the pulse repetition rate is increased tenfold, we observe a broadening of the first harmonic peaks, appearing at $\pm1.8$ T. The inhomogeneous spread of electronic $g$-factor $\Delta g$ induces a spread of spin precession frequencies $\Delta\omega_L=\Delta g\mu_BB/\hbar$ resulting in a faster dephasing of the spin ensemble, which is accelerated as the field becomes stronger \cite{abragam1961principles,yugova2012coherent}.

The rapid modulation of the excitation light with the PEM between the two signs of circular polarization ($\sigma^+$ and $\sigma^-$) at 50 kHz impedes the transfer of angular momentum from optically polarized electrons to nuclei. In order to achieve pumping of the nuclear spins, a quarter-wave plate ($\lambda/4$) was added before the PEM so that the laser beam was now modulated from $\sigma^+$ to $\pi$. Optical pumping of electron and sequentially nuclear spins is carried out during $\sigma^+$ excitation while it is absent when the light is linearly polarized. The difference between the intensity of reflected light due to the optical pumping $\Delta R$ is detected with the photodiode PD connected to a lock-in amplifier and synchronized with the PEM. The build-up of the nuclear spin polarization in time was measured by constantly recording the intensity of the reflected light in the arbitrary magnetic field in which we want to explore the nuclear spin dynamics. A shifting of the RSA peaks was observed as a result of the increasing nuclear field which enhances or diminishes, depending on its sign, the effect of the external field. In order to start from an unpolarized nuclear spin state, each measurement was preceded by a stage where the laser was blocked by a mechanical shutter and the magnetic field was set to zero for one minute. Since the nuclear spin-lattice relaxation time $T_1$ for this sample in the absence of optical pumping and at zero magnetic field is about 1 s \cite{mocek2017high}, one minute dark interval was sufficient to allow for complete nuclear spin depolarization.

\section{Experimental results and discussion}

\begin{figure}
\center{\includegraphics[width=1\linewidth]{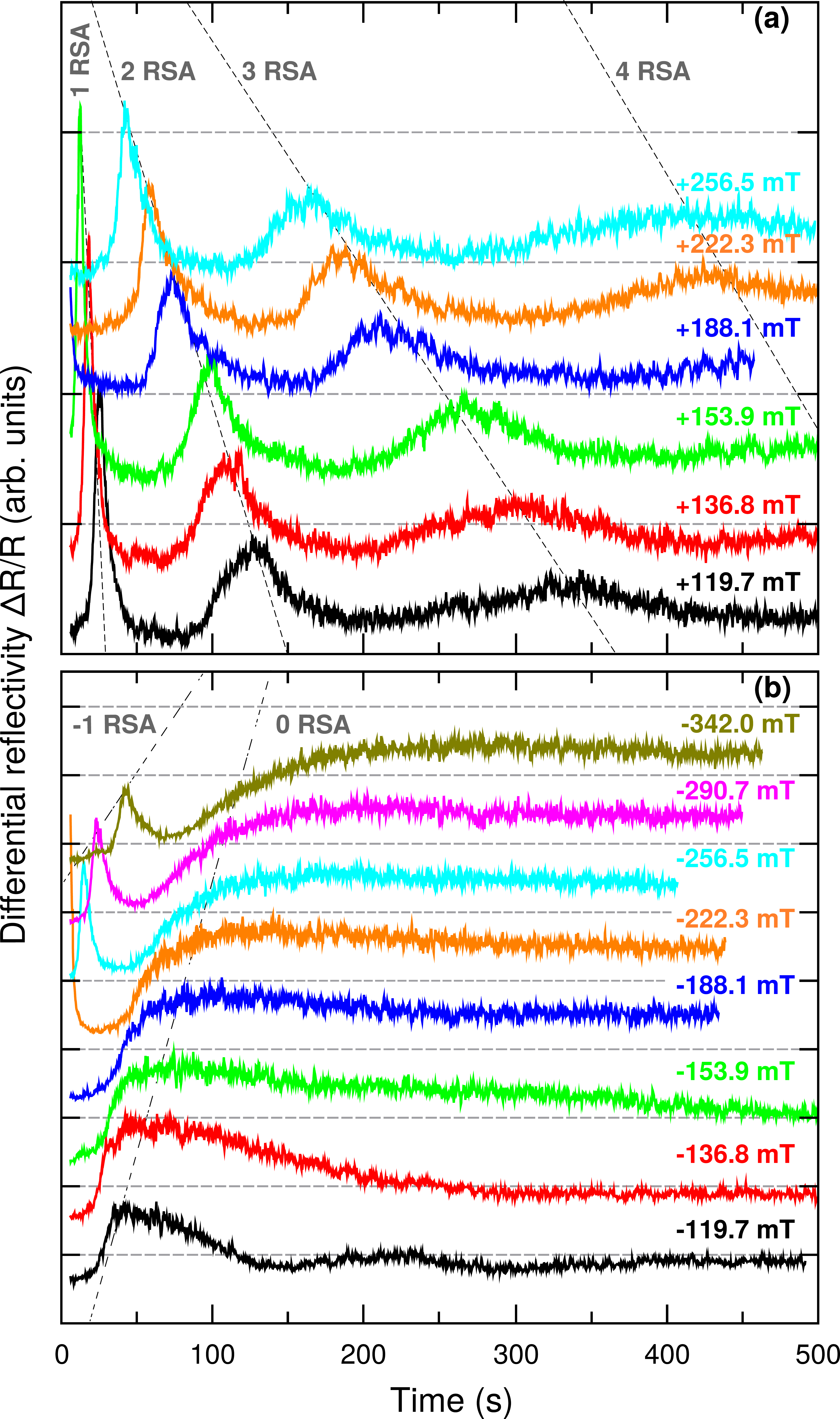}} 
\caption{Evolution of the differential reflectivity in time measured at $T=1.8\;\textnormal{K}$ for various (a) positive ($+B$) and (b) negative ($-B$) external magnetic fields. Laser repetition rate was $f=1\;\textnormal{GHz}$. Each curve is raised from the previous one for better visualization.}
\label{fig:PlusMinus}
\end{figure}

\begin{figure}
\center{\includegraphics[width=1\linewidth]{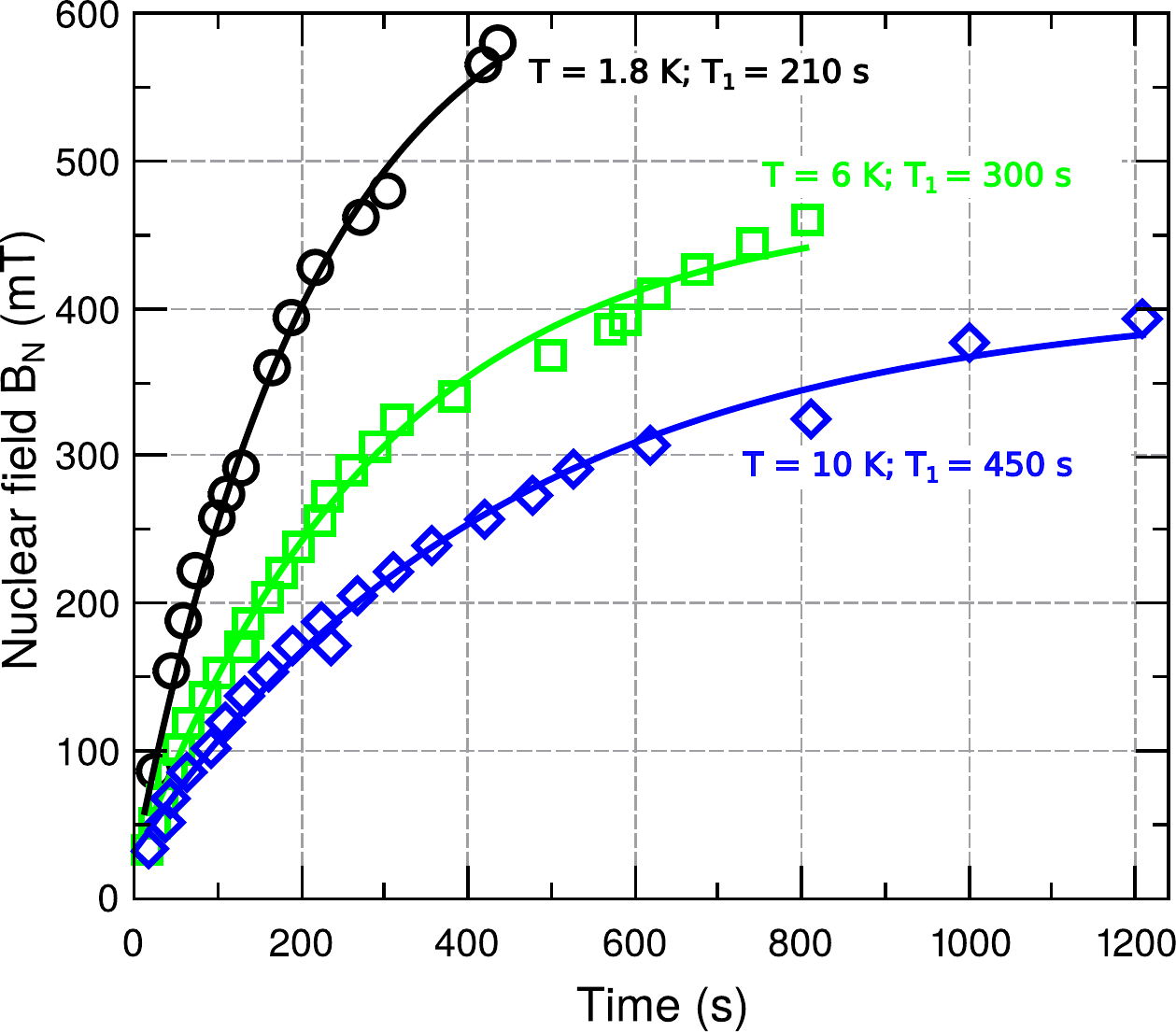}} 
\caption{Dynamics of the nuclear field measured for three different temperatures. Lines are fits with Eq. \eqref{eq:BN}.}
\label{fig:NuclearField}
\end{figure}

The evolution of the differential reflectivity for a set of different magnetic fields measured at $T=1.8\;\textnormal{K}$ is shown in Fig. \ref{fig:PlusMinus}. For both directions of the external magnetic field, we observe a shift of the RSA peaks related to the build-up of the nuclear polarization over time. When the external magnetic field is positive, i.e. parallel to the nuclear field, the electron spin is affected by the effective field equal to the sum of the external and nuclear field \citep{dyakonov2017spin}, $B_{eff}=B+B_N$. For a given external magnetic field, we observe a peak each time the resonance condition is fulfilled $\Omega_L=2\pi nf$, where $\Omega_L$ is the electron Larmor precession frequency in the effective magnetic field $B_{eff}$. Additionally, with the increase of the nuclear field, an amplitude decrease and a broadening of the peaks are perceived, originating from, most plausibly, an inhomogeneity of the nuclear field as a consequence of the spatial non-uniformity of the sample. An inhomogeneity of the excitation density of the laser spot could also cause a dispersion in the nuclear spin polarization dynamics, however, no significant change in the nuclear spin polarization time $T_1$ was observed when increasing the excitation light intensity. Since the exact positions of the RSA peaks in the absence of $B_N$ are known (Fig. \ref{fig:RSA}), it is possible to determine the current strength of the nuclear field from the maximum position of the corresponding peak for a certain external field ($B_N=B_{eff}-B$). In our previous measurements we observed no significant change in the nuclear spin polarization time when $B$ exceeded 50 mT \cite{mocek2017high}. Hence, it is possible to present the dependence of the nuclear field, determined for different external magnetic fields, on time for three different temperatures, see Fig. \ref{fig:NuclearField}. The change of the nuclear magnetic field with time can be expressed as \cite{meier2012optical}
\begin{equation}
B_N(t)=B_N^{st} \big[1-exp({-t/T_1})\big].
\label{eq:BN}
\end{equation}
The nuclear spin relaxation time $T_{1}$ has two contributions $1/T_1=1/T_{1e}+1/T_L$, where $T_{1e}$ is the relaxation time associated to the hyperfine interaction with electrons while $T_L$ is the relaxation time through alternative channels. Using the given equation to fit our experimental data it is possible to determine the nuclear spin polarization build-up time $T_1$ and the stationary value for the nuclear magnetic field $B_N^{st}$. The obtained values are listed in Table \ref{tb:Values}. 
\begin{table}[]
\caption{Experimentally determined nuclear spin polarization build-up time and stationary nuclear magnetic field for three different temperatures $T$.}
\begin{tabular}{ C{1.5cm}  C{1.5cm}  C{1.5cm} } \toprule
 $T\,(\textnormal{K})$ & $T_1\,(\textnormal{s})$ & $B_N^{st}\,(\textnormal{mT})$ \\ \hline
 1.8  & 210 & 629 \\
 6  & 300 & 447 \\
 10 & 450 & 375 \\
 \toprule
\end{tabular}
\label{tb:Values}
\end{table}
It can be clearly seen that with temperature increase the effectiveness of DNP drops. Nonetheless, the estimated nuclear spin polarization build-up time of 450 s at 10 K agrees well with the values measured on the same QW using the photoluminescence method for fields surpassing 50 mT and temperatures higher than 10 K, where it was found that the nuclei were polarized due to hyperfine scattering on itinerant electrons \cite{mocek2017high}. Temperature invariance of the time $T_1$ observed in Ref. [\onlinecite{mocek2017high}] is in agreement with the theoretical prediction for GaAs QW where the main relaxation channel is the hyperfine interaction with non-degenerate free electrons \cite{kalevich1990optical}
\begin{equation}
\frac{1}{T_{1e}}\propto\frac{A^2\Omega^2n_em^*}{\hbar^3d^2}.
\end{equation}
Here $A$ is the hyperfine constant, $\Omega$ is the unit cell volume, $n_e$ is the 2D electron concentration, $m^*$ is the effective mass of the electron and $d$ is the width of the QW. In this work the measurements were carried at temperatures lower than in Ref. [\onlinecite{mocek2017high}], therefore, a probable justification for the observed $T_1$ shortening with temperature can be related to electron localization at low temperatures. Localized electrons can transfer their spin to the nuclei more efficiently than free electrons, owing to the longer correlation time, resulting in a faster nuclear spin dynamics \cite{meier2012optical}. At higher temperatures the electrons become mobile due to thermal activation and an increase of $T_1$ time is observed.

When the external magnetic field $B$ is negative, the effective field $B_{eff}$ becomes smaller since the nuclear field $B_N$ compensates the external one. The change of the differential reflectivity in time for a given set of external magnetic fields due to the build-up of nuclear spin polarization is presented in Fig \ref{fig:PlusMinus}(b). Instead of observing several characteristic resonance peaks, "locking" of the nuclear field $B_N$ is detected, which becomes a constant after reaching the strength of the external magnetic field $B$. The reason for such behavior can be found in the anisotropy of the electron $g$-factor. Similar findings in a GaAs QW detected via the Hanle effect were reported in Refs. [\onlinecite{snelling1991magnetic}, \onlinecite{kalevich1992anisotropy}].

\begin{figure}[t]
\center{\includegraphics[width=1\linewidth]{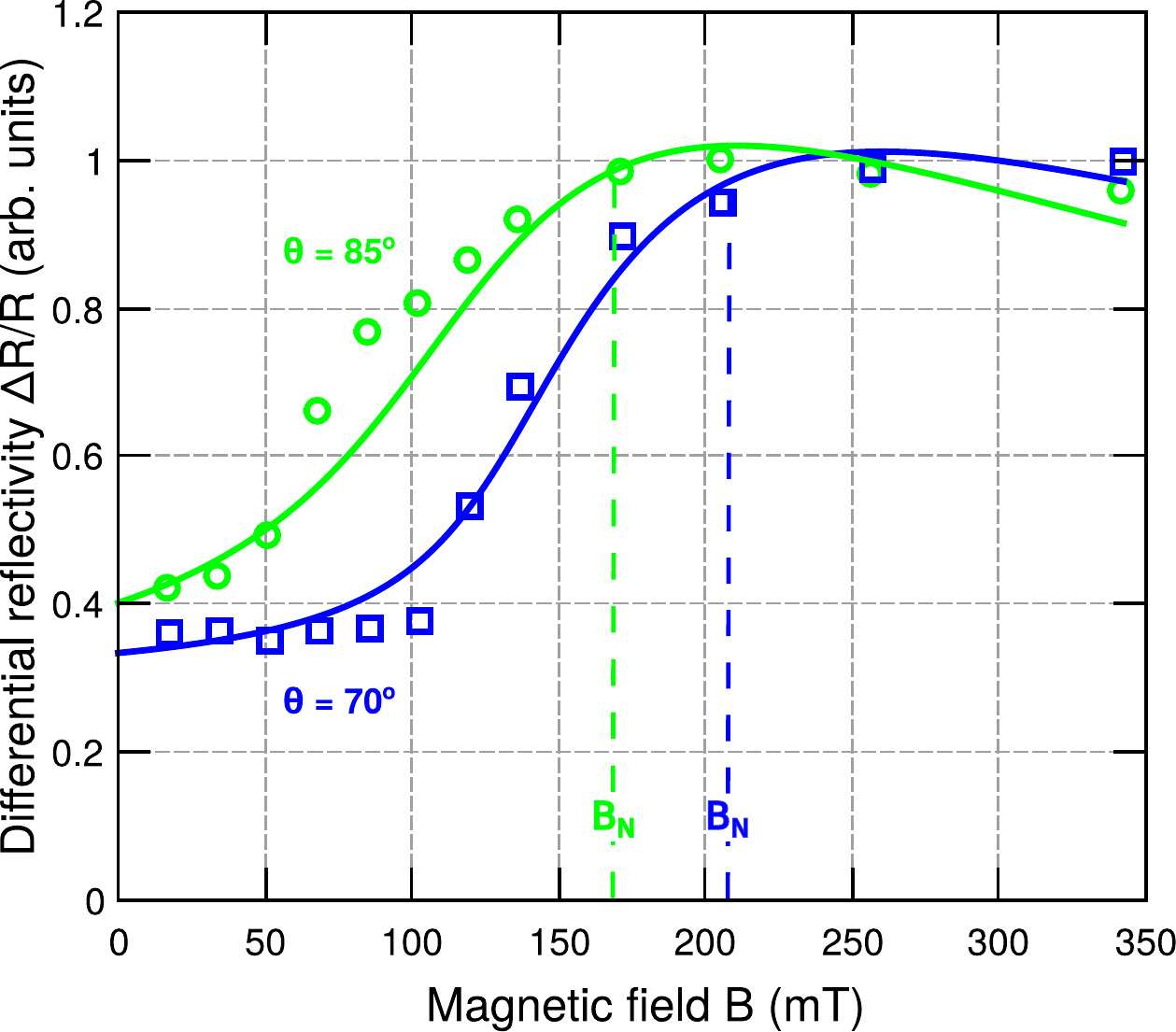}} 
\caption{Dependence of the differential reflectivity at saturation on magnetic field measured for two angles: $\theta = 85^{\circ}$ (green circles) and $\theta = 70^{\circ}$ (blue squares). Temperature is $T=1.8\;\textnormal{K}$ and laser with $f=10\;\textnormal{GHz}$ repetition rate is used. Lines show fits according to Eq. \eqref{eq:g-factor}.}
\label{fig:g-factor}
\end{figure}

In order to observe the shift of only the zero RSA peak we performed measurements with the 10 GHz laser where the external magnetic field was applied at an angle of $70\pm5^{\circ}$ or $85\pm5^{\circ}$ relative to the QW growth axis. Fig. \ref{fig:g-factor} illustrates how the value of differential reflectivity at saturation changes with the external magnetic field for the two angles. Considering that the measurements are done in oblique geometry, the electron-nuclear spin system becomes non-linear since the electron spin precession axis does not coincide with the direction of the external magnetic field. On the other hand, considering that the hyperfine interaction is isotropic, the nuclear field is collinear with the external one. Ergo, the direction of the precession axis of the electron spin is altered with variation of the strength of the total field $B_{eff}$. When the electron $g$-factor is isotropic the spin projection on the direction of light ($z$-axis) is given by $S_z = S_B\textnormal{cos}\,\theta = S_0\textnormal{cos}^2\,\theta$, where $S_0$ is the average electron spin in zero magnetic field, $\theta$ is the angle between the external magnetic field and the exciting light and $S_B = S_0\textnormal{cos}\,\theta$ is the spin projection along the magnetic field proportional to the nuclear field $B_N\propto S_B=S_0\textnormal{cos}\,\theta$. When lowering the angle from $85^{\circ}$ to $70^{\circ}$, $\textnormal{cos}\,\theta$ and consequently $B_N$ should increase fourfold. Even though we observe a small shift toward stronger magnetic fields, the difference between nuclear fields determined for the two angles is almost negligible (Fig. \ref{fig:g-factor}). In the interest of finding the ratio between the components of the $g$-factor parallel $g_{\parallel}$ and perpendicular $g_{\bot}$ to the growth axis we used the steady-state equation for the average electron spin $S$ when it is influenced by the external and nuclear fields \cite{kalevich1992anisotropy}
\begin{equation}
\boldsymbol{s}=\boldsymbol{s}_0+\bigg[\frac{g_{\bot}}{|g_{\bot}|}\boldsymbol{B}+\frac{g_{\parallel}-g_{\bot}}{|g_{\bot}|}(\boldsymbol{Bk})\boldsymbol{k}+a(\boldsymbol{sb})\boldsymbol{b}\bigg]\times \boldsymbol{s},
\label{eq:g-factor}
\end{equation}
where $\boldsymbol{s}=\boldsymbol{S}/|S_0|$, $\boldsymbol{s_0}=\boldsymbol{S_0}/|S_0|$, $\boldsymbol{k}$ is the unit vector along the $z$-axis, $b=\boldsymbol{B}/B$, and $a=5S_0A/\mu_BB_{1/2}|g_{\bot}|$ (here $B_{1/2}$ is the half-width of the Hanle curve, $A$ is the hyperfine constant and $\mu_B$ is the Bohr magneton, $A/\mu_B=1.55\,\textnormal{T}$ [\onlinecite{paget1977low}]). Fitting the numerical solution of the above equation to our experimental data, where the ratio between the two components of the $g$-factor was used as a fit parameter, yielded $g_{\bot}/ g_{\parallel}=1.3$. This value is in good agrement with the theoreticaly predicted one \cite{ivchenko1992electron}. The signs of the $g_{\bot}$ and $g_{\parallel}$ were taken from the Ref. [\onlinecite{ivchenko1992electron}] where for a 20 nm wide QW $g_{\parallel}<g_{\bot}<0$ should hold. For $B_{1/2}$ we used 7.5 mT, the half-width value of the zero RSA peak (Fig. \ref{fig:RSA}).

\section{Conclusions}

We have studied the nuclear spin dynamics in a nominally undoped GaAs/(Al,Ga)As QW by employing the method based on optical pumping of the resident electron spins in the rotating frame using high repetition periodic sequences from a Ti-Sapphire pulsed laser. From the resonance condition, when the Larmor precession frequency is synchronized with the pulse repetition rate, it was possible to obtain the value for the electron $g$-factor. The state of the electron-nuclear spin system is monitored through the changes in the intensity of the reflected light. This approach allowed us to follow the nuclear spin polarization accumulation in time and how it is affected by the change of the external magnetic field and temperature. It was found that the effectiveness of the dynamical nuclear polarization decreases with temperature. Also, "locking" of the nuclear field $B_N$ was observed which rises from zero to the value of the external magnetic field applied at that moment, without further increase. It was related to the anisotropy of the electron $g$-factor and the ratio between the in-plane $g_{\parallel}$ and out-of-plane $g_{\bot}$ components was determined.

\section*{Acknowledgments} We are grateful to K. V. Kavokin for useful discussions and to P. S. Sokolov for providing us with the algorithm used to determine the $g$-factor ratio. This work was supported by the Deutsche Forschungsgemeinschaft through the International Collaborative Research Centre 160 (Projects A6 and C7) and the DAAD (project EXCIPLAS). V.L.K. acknowledge the Russian Foundation for Basic Reasearch (Project No. 15-52-12017 NNIO a) for the financial support.

\bibliography{References}

\end{document}